\documentclass{ws-ijgmmp}

\begin{document}

\markboth{Shibendu Gupta Choudhury, Ananda Dasgupta, Narayan Banerjee}
{Raychaudhuri equation in scalar-tensor theory}

%
\catchline{}{}{}{}{}
%

\title{RAYCHAUDHURI EQUATION IN SCALAR-TENSOR THEORY}

\author{SHIBENDU GUPTA CHOUDHURY}

\address{Department of Physical Sciences, Indian Institute of Science Education and Research Kolkata, Mohanpur\\
Nadia, West Bengal 741246, India\\
\email{sgc14ip003@iiserkol.ac.in}}

\author{ANANDA DASGUPTA}

\address{Department of Physical Sciences, Indian Institute of Science Education and Research Kolkata, Mohanpur\\
Nadia, West Bengal 741246, India\\
\email{adg@iiserkol.ac.in}}

\author{NARAYAN BANERJEE}
\address{Department of Physical Sciences, Indian Institute of Science Education and Research Kolkata, Mohanpur\\
Nadia, West Bengal 741246, India\\
\email{narayan@iiserkol.ac.in}}

\maketitle

\begin{history}
\received{(Day Month Year)}
\revised{(Day Month Year)}
\end{history}

\begin{abstract}
Implications of the Raychaudhuri equation in focusing of geodesic congruences are studied in the framework of scalar-tensor theory of gravity. Specifically, we investigate the Brans-Dicke theory and Bekenstein's scalar field  theory. In both of these theories, we deal with a static spherically symmetric distribution and a spatially homogeneous and isotropic cosmological model as specific examples. We find that it is possible to violate the convergence condition under  reasonable physical assumptions. This leads to the  possibility of avoiding a singularity. 
\end{abstract}


\section{Introduction}
General Relativity (GR) is the most successful theory of gravity to date. It has enjoyed a great amount of success in describing physical reality. Still, there are a few drawbacks of GR. The inevitable existence of spacetime singularities in this theory is the biggest  concern among all of these. This generic existence of singularities has been proved in terms of the famous singularity theorems\cite{penrose, hawking}. Focusing of geodesic congruences is a central factor in proving these theorems. The Focusing Theorem (FT)\cite{poisson, wald} states that
 an initially converging timelike geodesic congruence will focus within a finite affine parameter value when the strong energy 
condition (SEC) holds. The SEC is a  reasonable physical assumption about the matter content of a spacetime. The FT is the most important consequence of the Raychaudhuri equation (RE)\cite{rc, ehlers}.
Now, the RE is a purely geometric equation. One has to invoke the Einstein equations, which relate the matter content of a 
system to the spacetime geometry, to arrive at the FT. It should be noted that focusing of a congruence leads to the formation of a congruence 
singularity - this  may or may not be a curvature singularity.  The concept of focusing along with a few additional reasonable conditions on spacetime enforces the general 
existence of curvature singularities\cite{penrose, hawking}. We refer to \cite{wald, hawell, joshi} for elegant and detailed discussions on the singularity theorems.

The
SEC, coupled with the Einstein equations, yields the convergence condition (CC) in GR.
The CC implies that gravity is attractive and this in turn leads to geodesic focusing. 
In other relativistic theories of gravity, the field equations are different. Thus, there is a possibility of violation of the CC, even when the  SEC is assumed.

 Burger {\it et al} studied the implications of the RE towards focusing of geodesic congruences in string theory,  braneworld gravity, $f(R)$ theory  and loop quantum  cosmology\cite{burger}. The RE has been applied in $f(R)$ gravity to study  the effective energy conditions and the possibility of obtaining repulsive 
gravity\cite{santos1, albareti, santos2}.

In the current  work, we consider Non-Minimally Coupled Scalar-Tensor Theories of gravity (NMCSTT).  Our aim is to use the RE to examine whether the CC is satisfied in these theories. Non-minimally coupled theories like the Brans-Dicke theory or $f(R)$ gravity theories have enjoyed a recent rejuvenation in the context of the accelerated expansion of the universe. For a very recent review on this, we refer to the work of Capozziello, D'Agostino and Luongo\cite{luongo1}. One should also note that in the absence of a universally accepted model for the accelerated universe, cosmographic quantities like the expansion scalar, acceleration etc are becoming more and more important in cosmology and this enhances the relevance of the RE in cosmology. We refer to \cite{luongo2, luongo3} for comprehensive reviews on cosmography. We start with the RE for geodesic motion for a general class of NMCSTT. Then we specialize to the Brans-Dicke theory\cite{brans1} and Bekenstein's scalar field 
theory\cite{bekenstein}. A large number of scalar field theories can be dealt with using  the general framework which is discussed here.   However, we will see that  in these theories, unlike in GR, it is very difficult  to reach a concrete conclusion without using some kind of an  exact solution. For the sake of definiteness, we shall consider examples from static spherically symmetric spacetimes and spatially isotropic and homogeneous cosmological spacetimes. The specific motivations for these choices will be mentioned in the corresponding sections.

This paper is organized as follows. In section \ref{sec2}, we briefly discuss the RE and the CC in the context of GR. We introduce NMCSTT in section \ref{sec3}. In this section, we derive the general expressions which  are necessary for the rest of the work. Section \ref{sec4} describes an examination of the CC for the Brans-Dicke and the Bekenstein scalar field theory 
in static spherically symmetric spacetimes. Section \ref{sec5} discusses  the same for cosmological spacetimes. The last section (section \ref{sec6}) consists of some concluding 
remarks.

\section{The RE and the CC in GR}\label{sec2}
The RE for a timelike geodesic congruence with velocity vector $u^\mu$ is given by\cite{rc, ehlers},
\begin{equation}\label{raych-eq0}
 \frac{\mathrm{d}\theta}{\mathrm{d}\tau}=-\frac{1}{3}\theta^2-\sigma_{\mu\nu}\sigma^{\mu\nu}
 +\omega_{\mu\nu}\omega^{\mu\nu}-R_{\mu\nu}u^\mu u^\nu,
 \end{equation}
where $\theta=\nabla_\mu u^\mu$ is the expansion scalar, $\tau$ is the proper time, 
$\sigma_{\mu\nu}=\nabla_{(\nu}u_{\mu)}-\frac{1}{3}h_{\mu\nu}\theta$
is the shear tensor with $h_{\mu\nu} = g_{\mu\nu} + u_{\mu}u_{\nu}$ being the spatial metric, $\omega_{\mu\nu}=\nabla_{[\nu}u_{\mu]}$ 
is the rotation tensor and $R_{\mu\nu}$ is the Ricci tensor.

We shall only consider hypersurface orthogonal congruences in this work. Therefore, we have $\omega_{\mu\nu}=0$ and Eq. \eqref{raych-eq0} becomes, 
\begin{equation}\label{raych-eq}
 \frac{\mathrm{d}\theta}{\mathrm{d}\tau}=-\frac{1}{3}\theta^2-\sigma_{\mu\nu}\sigma^{\mu\nu} -R_{\mu\nu}u^\mu u^\nu.
 \end{equation}
 It should be noted that both Eq. \eqref{raych-eq0} and  Eq. \eqref{raych-eq} are purely geometrical.  A theory of gravity comes into the picture when the field 
 equations are used to replace the term $R_{\mu\nu}u^\mu u^\nu$. 

 We assume that the matter distribution satisfies the SEC, i.e.,
 $T_{\mu\nu}u^\mu u^\nu+\frac{1}{2}T\geq 0.$
Let us now insert the Einstein equations,
$
 R_{\mu\nu}-\frac{1}{2} g_{\mu\nu} R=T_{\mu\nu},
$
into the SEC. It follows that
$
 R_{\mu\nu}u^\mu u^\nu\geq 0,
$
which is the so-called CC. With this condition, Eq. \eqref{raych-eq} implies,
$
 \frac{\mathrm{d}\theta}{\mathrm{d}\tau}+\frac{1}{3}\theta^2\leq 0.
$
Therefore, an initially converging timelike geodesic congruence develops a caustic ($\theta\rightarrow -\infty$) within finite proper time\cite{poisson,wald}. This is known as 
the FT.

With this brief recapitulation of the CC for GR, we will now proceed to examine the CC for NMCSTT.

\section{NMCSTT and the RE}\label{sec3}
A very general class of NMCSTT is given by the action,
\begin{equation}\label{action}
 S=\int \sqrt{-g}\mathrm{d}^4x\left[f(\phi)R-\frac{\omega(\phi)}{\phi}\partial_\mu\phi \partial^\mu\phi+2\mathcal{L}_m\right].
\end{equation}
Attempts towards such a general formulation of NMCSTT commenced long back, for instance by Nordtvedt\cite{nordt}.   While  $\omega$ is an 
arbitrary function of $\phi$ in Nordtvedt's formulation, $f(\phi)$ is chosen specifically to be equal to $\phi$.  Further generalizations along the same lines were  also discussed
by Wagoner\cite{wago}. Different scalar-tensor theories can be obtained  from the action (\ref{action}) for specific 
choices of $f(\phi)$ and $\omega (\phi)$. For a comprehensive discussion regarding scalar-tensor theories of gravity, we refer to \cite{fujii, quiros, faraoni1} and references 
therein.

We get the following field equations after varying the action \eqref{action} with respect to the metric ($g_{\mu\nu}$) and the scalar field ($\phi$), respectively - 

\begin{equation}\label{fsce1}
 f\left(R_{\mu\nu}-\frac{1}{2}g_{\mu\nu}R\right)+g_{\mu\nu}\Box f-\nabla_\nu \nabla_\mu f-\frac{\omega}{\phi}\partial_\mu\phi \partial_\nu\phi+\frac{1}{2}g_{\mu\nu}\frac{\omega}{\phi}\partial_\mu\phi \partial^\mu\phi=T_{\mu\nu},
\end{equation}

\begin{equation}\label{fsce2}
 R \frac{df}{d\phi}+\frac{2\omega}{\phi}\Box\phi+\left(\frac{1}{\phi}\frac{d\omega}{d\phi}-\frac{\omega}{\phi^2}\right)\partial_\mu\phi \partial^\mu\phi=0.
\end{equation}

We have from Eq. \eqref{fsce1},
\begin{equation}\label{maineq}
R_{\mu\nu}u^\mu u^\nu=\frac{1}{2f}\left[2(T_{\mu\nu}u^\mu u^\nu+\frac{1}{2}T)-\Box f +2 u^\mu u^\nu\nabla_\nu \nabla_\mu f +\frac{2\omega}{\phi}u^\mu u^\nu\partial_\mu\phi \partial_\nu\phi \right].
\end{equation}
It is clear from Eq. \eqref{maineq} that assuming the SEC does not necessarily imply the CC in NMCSTT.  Our aim is to explore the nature of the quantity 
$R_{\mu\nu}u^\mu u^\nu$ in these theories. In particular, we investigate the possibility of this quantity being negative despite the assumption of the SEC.

We shall use this equation (Eq. \eqref{maineq}) in two different theories of gravity, namely the Brans-Dicke theory\cite{brans1} and Bekenstein's conformally coupled scalar-tensor theory\cite{bekenstein}. The Brans-Dicke theory is perhaps the most talked about generalization of GR. The theory was believed to have GR as limiting case when $\omega$ goes to infinity\cite{weinberg}. This was later shown to be limited by the trace of the energy momentum tensor of the matter distribution\cite{soma, valerio2}. In spite of this, the Brans-Dicke theory is found to be extremely useful in resolving many cosmological problems. This includes, for example, the graceful exit problem in 
an inflationary paradigm\cite{johri, stein} and driving an accelerated expansion without any dark energy\cite{banerjee}. The motivation behind choosing our next example,  Bekenstein's scalar field, is that it has been widely discussed in the context of the No-Hair conjecture for black holes(\cite{boch}--\cite{her}). A technical application of Bekenstein's work is that it leads to a technique for generating solutions   for interacting conformally invariant models\cite{abreu}. 

Even for a specific theory, it is rarely possible to draw definite  conclusions  about the CC from the general equation (\ref{maineq}). Therefore, we shall pick up two representative exact solutions from both the theories for the 
actual detailed calculations. One is a static spherically symmetric solution which will help us understand the possibility of existence of black holes. The other is a spatially homogeneous and isotropic cosmological solution which will  let us investigate  the possibility of  avoiding a big bang type singularity. It should be emphasized that we have chosen isotropic cases both for the stationary and cosmological scenarios as these are the simplest examples. The purpose is to show how the equation (\ref{maineq}) can lead to some useful results. Other examples, like axially symmetric exact solutions may lead to different conclusions. However, the difference in these cases  will be caused  by the shear term ($\sigma_{\mu\nu}\sigma^{\mu\nu}$) or the vorticity term ($\omega_{\mu\nu}\omega^{\mu\nu}$) in equation (\ref{raych-eq0}) rather than specific characteristics of the scalar-tensor theory.

\section{Static spherically symmetric case}\label{sec4}
We begin by considering  a spherically symmetric static spacetime. We investigate the CC for the  two mentioned sub-classes of NMCSTT in this spacetime. 

\subsection{The Brans-Dicke theory}

The Brans-Dicke theory can be recovered as a special case from the action \eqref{action} when $\omega=\mbox{constant}$ and $f(\phi)=\phi$. The metric for a static spherically symmetric spacetime can be written in isotropic coordinates as, 
\begin{equation}\label{bdmetric}
\mathrm{d}s^2= -A^2(r)\mathrm{d}t^2+B^2(r)(\mathrm{d}r^2+r^2\mathrm{d}\theta^2+r^2\sin^2\theta \mathrm{d}\varphi^2).
\end{equation}

We choose a freely falling observer,
\begin{equation}
 u^\alpha=\left(\frac{1}{A^2},-\frac{\sqrt{1-A^2}}{AB},0,0\right).
\end{equation}
Here, the  negative radial component indicates that we are considering  incoming geodesics. 
Using this, we have from Eq. \eqref{maineq},
\begin{equation}\label{bdstatic}
\begin{split}
 R_{\mu\nu}u^\mu u^\nu=\frac{1}{2\phi}\left[2\left(T_{\mu\nu}u^\mu u^\nu+\frac{1}{2}T\right)+\frac{2-3A^2}{A^2}\Box\phi+
 \frac{2\omega}{\phi}\frac{1-A^2}{A^2B^2}(\phi^\prime)^2\right.\\ \left.-\frac{2\phi^\prime}{A^2 B^2}\left({2(1-A^2)}\frac{B^\prime}{B}+(2-A^2)\frac{A^\prime}{A}+
 \frac{2}{r}{(1-A^2)}\right)\right].
 \end{split}
\end{equation}

A combination of Eq. \eqref{fsce2} and Eq. \eqref{bdstatic} obtains,
\begin{equation}\label{rcbdsss}
\begin{split}
 R_{\mu\nu}u^\mu u^\nu=P_1+P_2+P_3+P_4,
 \end{split}
\end{equation}
where \begin{equation}\label{p1234}
\begin{split}
 P_1 &=\frac{1}{\phi}\left(T_{\mu\nu}u^\mu u^\nu+\frac{1}{2}T\right),\\
 P_2 &=\frac{3A^2-2}{A^2}\frac{R}{4\omega},\\ P_3 &=\frac{(\phi^\prime)^2}{2 A^2B^2\phi^2}\left((1+2\omega)-\left(\frac{3}{2}+2\omega\right)A^2\right),\\P_4 &= -\frac{\phi^\prime}{A^2 B^2\phi}\left({2(1-A^2)}\frac{B^\prime}{B}+(2-A^2)\frac{A^\prime}{A}+\frac{2}{r}{(1-A^2)}\right).
 \end{split}
\end{equation}
Here we have split up the right hand side of Eq. \eqref{rcbdsss} into  four terms as indicated for the sake of convenience of discussion. We now discuss these terms one by one.

We demand a positive $f(\phi)=\phi$ in order to ensure a positive gravitational coupling. Therefore, $P_1$ is always non-negative as long as the SEC is obeyed. $P_2$ switches its sign 
at $A^2=\frac{2}{3}$, whereas $P_3$ does so at $A^2=\frac{2+4\omega}{3+4\omega}$. It follows that we may have terms with negative contribution to $R_{\mu\nu}u^\mu u^\nu$. The latter can  be negative if these terms dominate even if the SEC is satisfied. Therefore, the CC will be violated in such a case.

 We consider the  Brans Class I solution\cite{brans1,
thesis, brans2} as an explicit example. This solution has been widely discussed in the literature for its usefulness in the context of local astronomical tests and the boundary condition problem\cite{brans1,bhadra,janis, camp}. 

Considering the weak field limit and assuming asymptotic flatness, the Brans Class I solution is given by Eq. \eqref{bdmetric} with\cite{brans1},
\begin{eqnarray}
A^2&=&\left(\frac{1-\frac{D}{r}}{1+\frac{D}{r}}\right)^{\frac{2}{\lambda}}, \\
B^2&=&\left(1+\frac{D}{r}\right)^4\left(\frac{1-\frac{D}{r}}{1+\frac{D}{r}}\right)^{\frac{2(\lambda-C-1)}{\lambda}},\\
\phi&=& \frac{1}{\lambda^2 G_0}\left(\frac{1-\frac{D}{r}}{1+\frac{D}{r}}\right)^{\frac{C}{\lambda}}. 
\end{eqnarray}
Here,
 \begin{eqnarray}
   \lambda=\sqrt{\frac{2\omega+3}{2\omega+4}}, \hspace{0.2cm} C=-\frac{1}{2+\omega}, \hspace{0.2cm} D=\frac{MG_0\lambda}{2}.
 \end{eqnarray}
$M$ is identified as mass of the static spherical distribution and $G_0$ is the Newtonian gravitational coupling constant. This 
solution is valid for $\omega>-\frac{3}{2}$ and $\omega<-2$.\\

Since the Brans class I solution is a vacuum solution, we have $P_1=0$. The remaining terms in Eq. \eqref{rcbdsss} are given by,
 \begin{eqnarray}
   P_2 &=&\left(\frac{3A^2-2}{A^2}\right)\frac{D^2 C^2}{r^4\lambda^2 \left(1-\frac{D^2}{r^2}\right)^4 }\left(\frac{1-\frac{D}{r}}{1+\frac{D}{r}}\right)^{\frac{2(C+1)}{\lambda}},\label{p2ex}\\
 P_3 &=& \frac{(\phi^\prime)^2}{2A^2B^2\phi^2}\left((1+2\omega)-\left(\frac{3}{2}+2\omega\right)A^2\right), \label{p3ex}\\ \label{P4eq}
 P_4 &=& \frac{1}{A^2B^2}\left(\frac{8(1-A^2)D^2 C(C r-D \lambda +r)}{\lambda ^2 r^5 \left(1-\frac{D^2}{r^2}\right)^2}\right. -\frac{4(2-A^2) D^2 C}{\lambda ^2 r^4 \left(1-\frac{D^2}{r^2}\right)^2}\\ \nonumber && \hspace{1.5cm} \left.-\frac{4{(1-A^2)} D C}{r^3\lambda\left(1 -\frac{D^2}  {r^2}\right)}\right).
 \end{eqnarray}
One can verify from Eq. \eqref{p2ex}  that $P_2$ is negative when $A^2<\frac{2}{3}$. Similarly, $P_3$ is negative when $\frac{2+4\omega}{3+4\omega}<A^2<1$. Understanding the sign of $P_4$ takes a bit more effort. For $C<0$ (i.e. $\omega>-2$), the first term in Eq. \eqref{P4eq} is negative when $r(1+C)>{D\lambda}$, whereas the other two terms are always positive. For 
$C>0$ (i.e. $\omega<-2$), the first term in Eq. \eqref{P4eq}  is negative when $r(1+C)<{D\lambda}$. The other two terms are always  negative in this case.\\

\begin{figure}[h]
\frame{\centerline{\psfig{file=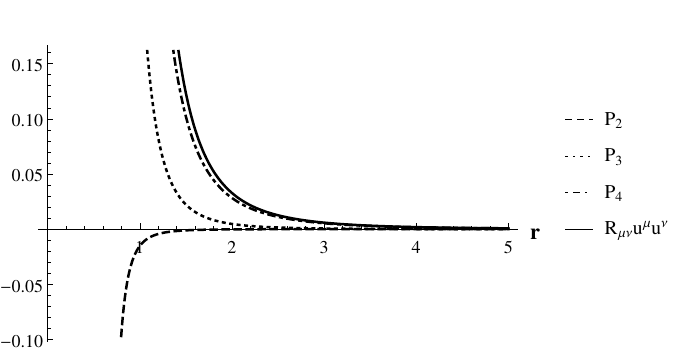,width=5in}}}
\vspace*{8pt}
\caption{{ Radial variation of different terms present in the expression of $R_{\mu\nu}u^\mu u^\nu$ (Eq. \eqref{rcbdsss}) for Brans Class I solution. We have chosen 
$G_0=1$, $M=1$ and a positive $\omega(=8)$.}
 \label{fig1}} 
\end{figure}

\begin{figure}[h]
\frame{\centerline{\psfig{file=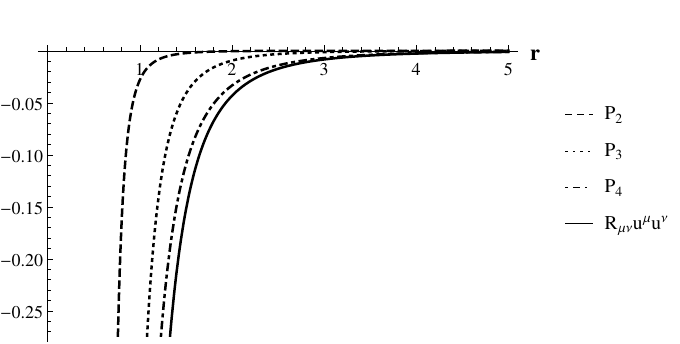,width=5in}}}
\vspace*{8pt}
 \caption{{ Radial variation of different terms present in the expression of $R_{\mu\nu}u^\mu u^\nu$ (Eq. \eqref{rcbdsss}) for Brans Class I solution. We have chosen 
 $G_0=1$, $M=1$ and a negative $\omega(=-8)$.}
 \label{fig2}} 
\end{figure}

To get a definite idea about the overall sign of $R_{\mu\nu}u^\mu u^\nu$,  we will now plot the behavior of the terms $P_2$,  $P_3$, $P_4$ and $R_{\mu\nu}u^\mu u^\nu$.
The plots for a positive and a negative value of $\omega$ are presented in 
Figs.  \ref{fig1} and \ref{fig2}, respectively. The qualitative  nature of the plots does  not appear to depend on $|\omega |$  as long as the sign of $\omega$ is fixed.
  
It is evident from the plots that $R_{\mu\nu}u^\mu u^\nu$ is negative only for allowed negative values of $\omega$ (Fig. \ref{fig2}). For positive $\omega$, 
$R_{\mu\nu}u^\mu u^\nu$ is always positive (Fig. \ref{fig1}). It approaches zero for large $r$ in both the cases. Therefore, for it to be possible to avoid a singularity in this specific model, $\omega$ has to be negative.

\subsection{Bekenstein's scalar field} 

The action in the case of Bekenstein's conformally coupled scalar-tensor theory  is given by, 
\begin{equation}
 S=\int \sqrt{-g}\mathrm{d}^4x\left[R-\partial_\mu\phi \partial^\mu\phi-\frac{1}{6}R\phi^2\right].
\end{equation}
Thus, we have $f(\phi)=1-\frac{\phi^2}{6}$ and $\omega(\phi)=\phi$.
{In this setting, two types of static spherically symmetric solutions were discussed in \cite{bekenstein}.}
Among these, we will consider the type A solution\cite{bekenstein} in this work. This solution is given by,
\begin{equation}\label{bks}
 \mathrm{d}s^2= - A^{2}(r) \mathrm{d}t^2+ B^{2}(r) \mathrm{d}r^2+ S^{2}(r) \left(r^2\mathrm{d}\theta^2+r^2\sin^2\theta \mathrm{d}\varphi^2\right),
\end{equation}
where,

\begin{equation}\label{abcp}
 \begin{split}
  A^2(r)=\frac{1}{4}\left[w(r)^\beta+ w(r)^{-\beta}\right]^{2}w(r)^{2\alpha},\\
  B^2(r)=\frac{1}{4}\left[w(r)^\beta+ w(r)^{-\beta}\right]^{2}w(r)^{-2\alpha},\\
  S^2(r)=\frac{1}{4}\left[w(r)^\beta+ w(r)^{-\beta}\right]^{2}w(r)^{-2\alpha+2}.
 \end{split}
\end{equation}

In Eq. \eqref{abcp}, $\alpha=\sqrt{1-3\beta^2}$ and $\beta$ is a constant in the range $-\frac{1}{\sqrt{3}}\leq\beta\leq \frac{1}{\sqrt{3}}$. The solution for the scalar field is,
\begin{equation}\label{conphi}
\phi = \sqrt{6} \frac{1 - w^{2\beta}}{1 + w^{2\beta}}.
\end {equation}
This solution \eqref{abcp} reduces to the Schwarzschild solution if we choose,
\begin{equation}\label{wr}
 w=\sqrt{1-\frac{2M}{r}},
\end{equation}
  and $\beta=0$\cite{bekenstein}. This is one of the reasons for choosing  this example. The solution \eqref{abcp} represents a black hole with a scalar hair for 
$\beta=\pm \frac{1}{2}$. This black hole solution is known as the Bocharova–Bronnikov–Melnikov–Bekenstein (BBMB) 
black hole\cite{beken3}. The BBMB black hole is the first known counterexample to the No-Hair conjecture on black holes. Xanthopoulos and Zannias\cite{xan} shown that this black hole solution is the unique static, asymptotically flat solution for the Einstein conformal-scalar system. The significance of this solution in the context of the No-Hair conjecture can be found in the references \cite{boch}--\cite{her}. Another important significance of the BBMB black hole is that with a negative scalar charge, it can mimic a wormhole or an Einstein-Rosen bridge (see \cite{avijit} and references therein). The metric \eqref{bks}  represents a naked singularity at $r=2M$ for all 
other values of $\beta$. The naked singularity is a 2-surface, except over the interval
$\frac{1}{2}<|\beta|\leq\frac{1}{\sqrt{3}}$. 

For a freely falling observer,
\begin{equation}
 u^\alpha=\left(\frac{1}{A^2},-\frac{\sqrt{1-A^2}}{AB},0,0\right),
\end{equation}
Eq. \eqref{maineq} yields,
\begin{equation}\label{rmunubb}
 \begin{split}
  R_{\mu\nu}u^\mu u^\nu=Q_1+Q_2+Q_3,
 \end{split}
\end{equation}
where
\begin{equation}
\begin{split}
Q_1 &=\frac{\phi\Box \phi}{\left(1-\frac{\phi^2}{6}\right)}\left(\frac{3A^2-2}{6A^2}\right),\\
Q_2 &=\frac{\phi\phi^\prime}{3A^2B^2{\left(1-\frac{\phi^2}{6}\right)}}\left[(2-A^2)\frac{A^\prime}{A}+2(1-A^2)\frac{S^\prime}{S}+\frac{2}{r}(1-A^2)\right],\\
Q_3 &= \frac{(\phi^\prime)^2}{{\left(1-\frac{\phi^2}{6}\right)}}\frac{4-3A^2}{6A^2 B^2}.
\end{split}
\end{equation}

The stress energy tensor of this scalar field is trace-free. Thus, the field equations lead to a zero Ricci scalar. The wave equation \eqref{fsce2} then leads to,
\begin{equation}
 \Box \phi=0 \implies Q_1=0.
\end{equation}
$Q_3$ is always  positive, as $A^2<1$ and $f(\phi)>0$. However, such general statements can not be made about the sign of $Q_2$ and hence that of $R_{\mu\nu}u^\mu u^\nu$. For this, 
we have to use explicit expressions for $A$, $B$, $S$ and $\phi$.

\begin{figure}[h]
\frame{\centerline{\psfig{file=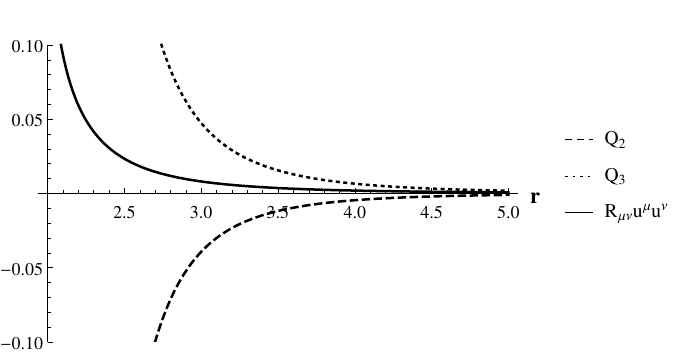,width=5in}}}
\vspace*{8pt}
 \caption{{Radial variation of different terms present in the expression \eqref{rmunubb} with  $M=1$ and $|\beta|=0.5$.}
 \label{fig3}} 
\end{figure}

\begin{figure}[h]
\frame{\centerline{\psfig{file=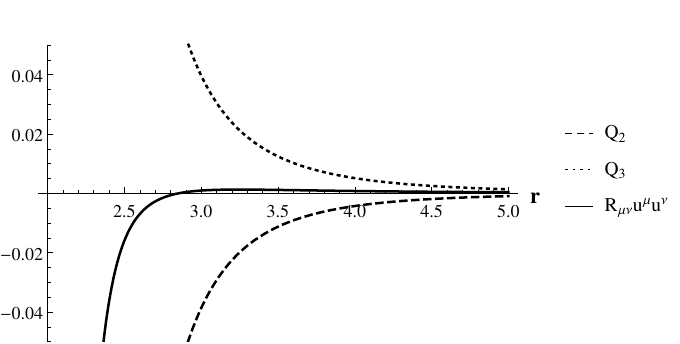,width=5in}}}
\vspace*{8pt}
 \caption{{Radial variation of different terms present in the expression \eqref{rmunubb} with $M=1$ and $|\beta|=0.4$.}
 \label{fig4}} 
\end{figure}

\begin{figure}[h]
\frame{\centerline{\psfig{file=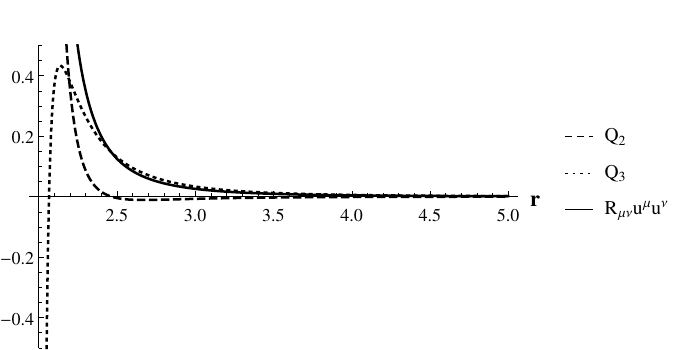,width=5in}}}
\vspace*{8pt}
 \caption{{ Radial variation of different terms present in the expression \eqref{rmunubb} with  $M=1$ and $|\beta|=0.57$.}
 \label{fig5}} 
\end{figure}

We will now use the expressions for $A$, $B$, $S$ and $\phi$, written in Eqs. \eqref{abcp} and \eqref{conphi} where we choose $w(r)$ as in the Eq. \eqref{wr}.
Using this, we plot the variation of $Q_2$, $Q_3$ and $R_{\mu\nu} u^\mu u^\nu$ with $r$. We present the plots for 
$|\beta|=\frac{1}{2}, \hspace{0.2cm} 0.4$ and $0.57$ in Figs. \ref{fig3}, \ref{fig4} and \ref{fig5}, respectively. These three different values of $|\beta|$ represent 
all possible distinctive features of the solution \eqref{abcp}.

We have found that for $|\beta|<\frac{1}{2}$, $R_{\mu\nu} u^\mu u^\nu $ is negative in a region near $r=2M$. With increasing $r$, it crosses zero, attains a maximum value  
and approaches zero for large $r$ (as shown in Fig. \ref{fig4}). For $\frac{1}{2}\leq|\beta|\leq\frac{1}{\sqrt{3}}$, $R_{\mu\nu} u^\mu u^\nu$ is always positive (as shown 
in Figs. \ref{fig3}, \ref{fig5}). Thus, violation of the CC is possible only for   $|\beta|<\frac{1}{2}$. This violation is restricted to a small range of $r$. This 
conclusion  is also in agreement with the results, discussed in \cite{bekenstein}.  For all other choices of $\beta$, the CC is always satisfied and a singularity is inevitable.

\section{Spatially homogeneous and isotropic case}\label{sec5}
In this section, we will  work  with a spatially homogeneous and isotropic spacetime - namely the Friedmann-Robertson-Walker (FRW) universe,
\begin{equation}
 \mathrm{d}s^2=-\mathrm{d}t^2+a^2(t)\left(\frac{\mathrm{d}r^2}{1-kr^2}+r^2\mathrm{d}\theta^2+r^2\sin^2\theta \mathrm{d}\varphi^2\right).
\end{equation}
We assume that the matter content of the universe consists of a perfect fluid, i.e.,
\begin{equation}
 T_{\mu\nu}=(\rho+p)u_\mu u_\nu+pg_{\mu\nu}.
\end{equation}
Here $u^\mu=(1,0,0,0)$.

We will again consider the cases  of the Brans-Dicke scalar field and the Bekenstein scalar field to examine the CC in this cosmological context.

\subsection{The Brans-Dicke theory}
For the Brans-Dicke scalar field, Eq. \eqref{maineq} implies,
\begin{equation}\label{frbd1}
 R_{\mu\nu}u^\mu u^\nu=\frac{\rho+3p}{2\phi}+\frac{3}{2}\frac{\ddot{\phi}}{\phi}+\frac{3}{2}\frac{\dot{a}}{a}\frac{\dot{\phi}}{\phi}+\omega\frac{\dot{\phi}^2}{\phi^2}.
\end{equation}
After some manipulations employing Eqs. \eqref{fsce1} and \eqref{fsce2}, we have,
\begin{equation}\label{frbd}
 R_{\mu\nu}u^\mu u^\nu=\frac{\omega}{(2\omega+3)\phi}(3p-\rho)+3\frac{\dot{a}^2}{a^2}+\frac{3k}{a^2}+\frac{\omega}{2}\frac{\dot{\phi}^2}{\phi^2}.
\end{equation}
 The second term in the right hand side of the above expression \eqref{frbd} is always positive. The third term is  negative for $k=-1$. The last term is negative when $\omega$ is negative. If $\rho>3p$, the first term is negative when $\omega>0$ or when $\omega<-\frac{3}{2}$. On the other hand, if 
$\rho<3p$, this term is negative for $-\frac{3}{2}<\omega<0$. 

It is clear from Eq. \eqref{frbd1} that the signs of the derivatives of $\phi$ and that of $\omega$ will play a crucial role in the CC. In order to have a general idea in the absence of exact solutions, we have used the field equations and arrived at Eq. \eqref{frbd}. The latter indicates that the spatial curvature $k$ plays a role in the CC. An open 
universe ($k=-1$) in the Brans-Dicke theory may actually lead to a non-singular model. We need an exact solution to explicitly determine the sign of $R_{\mu\nu}u^\mu u^\nu$.

We take the example of an exact solution for a spatially flat ($k=0$) geometry and with  pressure-less ($p=0$) matter. This solution was proposed by Brans and Dicke in their seminal paper \cite{brans1} and is given by,
\begin{equation}
 a=a_0 t^{\frac{2(\omega+1)}{3\omega+4}},
\end{equation}

\begin{equation}
 \phi=\frac{(3\omega+4)\rho_0 t^{\frac{2}{3\omega+4}}}{2a_0^3(2\omega+3)}.
\end{equation}

For this solution, we have,
\begin{equation}
 R_{\mu\nu}u^\mu u^\nu=\frac{6 (\omega +1) (\omega +2)}{t^2 (3 \omega +4)^2}.
\end{equation}
The above expression is negative for $-2<\omega< -\frac{3}{2}$. This is exactly the condition for accelerated expansion of the Universe, represented by this solution\cite{banerjee}. 
It is to be noted that violation of the CC in this case corresponds to observationally  unfavorable values of $\omega$\cite{brans1}. 

This is a very simple solution, but is sufficiently useful to explain that the CC does not follow  solely from the  SEC in the Brans-Dicke theory. The CC depends crucially on the value and sign of the 
parameter $\omega$ .

\subsection{Bekenstein's scalar field}
For the case of the Bekenstein scalar field, Eq. \eqref{maineq} obtains,
\begin{equation}
 R_{\mu\nu}u^\mu u^\nu=\frac{1}{2\left(1-\frac{\phi^2}{6}\right)}\left[(\rho+3p)+\dot{\phi}^2-\phi\ddot{\phi}-\frac{\dot{a}}{a}\phi\dot{\phi}\right].
\end{equation}
Using the field equations \eqref{fsce1} and \eqref{fsce2}, one can arrive at the following expression:
\begin{equation}\label{frbk}
 R_{\mu\nu}u^\mu u^\nu=\frac{3p-\rho}{2}+3\frac{\dot{a}^2}{a^2}+\frac{3k}{a^2}.
\end{equation}
The first term in the right hand side of the above Eq. \eqref{frbk} gives a negative contribution when $\rho>3p$. The third term is negative for $k=-1$. The second term is 
always positive. { Thus, with a radiation distribution ($p = \frac{1}{3} \rho$),  $R_{\mu\nu}u^\mu u^\nu$ can be negative only for an open universe ($k=-1$).}

Let us now consider explicit examples of exact solutions containing radiation for $k=0$, $1$ and $-1$. These solutions were all found by Bekenstein\cite{bekenstein}.

\begin{itemize}
 \item For the $k=0$ case,
\begin{equation}
 a(t)=C (t-t_0)^\frac{1}{2},
\end{equation}
where $C$ and $t_0$ are arbitrary constants.
Therefore, we have,
\begin{equation}
  R_{\mu\nu}u^\mu u^\nu=\frac{3}{4}\frac{1}{(t-t_0)^2},
\end{equation}
which is always positive. 

\item For the $k=1$ case,
\begin{equation}
 a=\sqrt{C^2-(t-t_0)^2},
\end{equation}
and
\begin{equation}
 R_{\mu\nu}u^\mu u^\nu=\frac{3C^2}{\left[C^2-(t-t_0)^2\right]^2},
\end{equation}
which is again positive definite. {Hence focusing of geodesics is inevitable for both $k=0$ and $k=1$. }

\item For $k=-1$, the solution is,
\begin{equation}\label{keqm1}
 a=\sqrt{C^2+(t-t_0)^2}.
\end{equation}
Here we have,
\begin{equation}
 R_{\mu\nu}u^\mu u^\nu=-\frac{3C^2}{\left[C^2+(t-t_0)^2\right]^2}.
\end{equation}
Thus, in this case, we have a possibility of defocusing of geodesics and one can avoid a singularity of  zero proper volume. 
\end{itemize}

These conclusions are in excellent agreement with those, obtained from Eq. \eqref{frbk}. The CC is violated for the $k=-1$ case only, and Bekenstein showed that 
the solution for the $k=-1$ case (Eq. \eqref{keqm1}) is indeed non-singular\cite{bekenstein}. It is well known that the observed universe appears to be spatially flat ($k=0$). So the $k=\pm 1$ cases are not really meant for describing observational cosmology. This is rather an exercise to check what spatial geometry could have possibly avoided the singularity in Bekenstein's scalar field theory.

\section{Conclusion}\label{sec6}
 
 The proofs of the singularity theorems\cite{penrose, hawking} rely on the focusing of geodesic congruences. The condition for focusing, namely the CC 
  is determined on the other hand by the RE\cite{rc, ehlers}. In GR, this condition is found to be satisfied for geodesics once we impose the
 quite reasonable SEC on the matter distribution. Thus a singularity appears to be inevitable in a purely gravitational 
 system in GR. For other relativistic theories of gravity, some characteristics of the theory may change the CC. In the present work, this modified 
 CC is investigated in two NMCSTT, which are the Brans-Dicke theory and Bekenstein's conformally coupled 
 scalar-tensor theory.
 
 It is found that the SEC alone cannot guarantee the CC in these cases, as opposed to the situation in GR. In order to arrive at definite  conclusions,  we chose some exact solutions for two representative situations in both the theories. One of them is the static spherically symmetric solution, which is the analogue  of the Schwarzschild metric. The second example corresponds to the Friedmann cosmologies.
 
 For the static spherical distributions, it is found that corresponding to a negative $\omega$ in the Brans-Dicke theory, it is possible to violate the CC even when the SEC holds. Similar conclusion follows for $|\beta| < \frac{1}{2}$ in Bekenstein's 
 theory. It deserves mention that while $\omega$ is a parameter in the Brans-Dicke theory, $\beta$ is an arbitrary constant  of integration (limited 
 to the range $-\frac{1}{\sqrt{3}}\leq\beta\leq \frac{1}{\sqrt{3}}$).
 
 In the context of cosmology, we conclude that at least for an open universe ($k=-1$), the CC can indeed be violated and hence a cosmological model without 
 a singularity may be obtained. While this is possible only for an open universe in Bekenstein's theory, we see that in the Brans-Dicke theory, violation of the CC may be achieved even 
 for a spatially flat universe. 
 
Similar investigations can be 
 carried out  for any specific NMCSTT which can be written as a special case of the action \eqref{action}. One should also note that these theories escape the purview of the singularity theorems as in references \cite{penrose} and \cite{hawking} which assume GR
 at the outset. However, since for every functional form of $\omega (\phi)$ and $f(\phi)$, one has a different theory of gravity, it will not be possible to generalize 
 the theorems for the generic action \eqref{action}.

\section*{Acknowledgments}
The authors thank the anonymous referee whose comments and suggestions improved the quality of the paper.
Shibendu Gupta Choudhury thanks Council of Scientific and Industrial Research, India for the financial support.


\begin{thebibliography}{99}

\bibitem{penrose} R. Penrose, Gravitational collapse and space-time singularities, {\it Phys. Rev. Lett.} {\bf 14} (1965),  57--59.

\bibitem{hawking} S. W. Hawking and R. Penrose, The Singularities of gravitational collapse and cosmology, {\it Proc. Roy. Soc. Lond. A} {\bf 314} (1970), 529--548.



\bibitem{rc} A. K. Raychaudhuri, Relativistic cosmology. I, {\it Phys. Rev.}  {\bf 98} (1955), 1123--1126.

\bibitem{ehlers} J. Ehlers, Beitr\"age zur relativistischen Mechanik kontinuierlicher Medien,  {\it Akad. Wiss. Lit. Mainz, Abhandl. Math.-Nat. Kl.} {\bf 11} (1961), 793--837; English translation: J. Ehlers, Contributions to the relativistic mechanics of continuous media, {\it Gen. Relativ. Gravit.} {\bf 25} (1993), 1225--1266.

\bibitem{poisson} E. Poisson, {\it A Relativist's Toolkit: The Mathematics of Black-Hole Mechanics} (Cambridge University Press, Cambridge 2004).

\bibitem{wald} R. M. Wald, {\it General Relativity} (Chicago University Press, Chicago 1984).

\bibitem{hawell} S. Hawking and G. Ellis, {\it The Large Scale Structure of Space-Time} 
(Cambridge University Press, Cambridge 2011).

\bibitem{joshi} P. Joshi, {\it Global Aspects in Gravitation and Cosmology},  International series of monographs on physics {\bf 87} (Oxford University Press, Oxford 1997).



\bibitem{burger} D. J. Burger, N. Moynihan, S. Das, S. S. Haque, and B. Underwood, Towards the Raychaudhuri Equation Beyond General Relativity, {\it Phys. Rev. D} {\bf 98} (2018), 024006.

\bibitem{santos1} J. Santos, J. S. Alcaniz, M. J. Reboucas and F. C. Carvalho, Energy conditions in $f(R)$ gravity,
{\it Phys. Rev. D}  {\bf 76} (2007), 083513.

\bibitem{albareti} F. D. Albareti, J. A. R. Cembranos, A. de la Cruz-Dombriz and
A. Dobado, On the non-attractive character of gravity in $f(R)$ theories, {\it J. Cosmology Astropart. Phys.} {\bf 1307} (2013), 009.

\bibitem{santos2} C. S. Santos, J. Santos, S. Capozziello and J. S. Alcaniz, Strong energy condition and the repulsive character of $f(R)$ gravity, {\it Gen.
Relativ. Gravit.} {\bf 49} (2017), 50.

\bibitem{luongo1} S. Capozziello, R. D'Agostino and O. Luongo, Extended gravity cosmography, {\it Int. J. Mod. Phys. D} {\bf 28} (2019) 10, 1930016. 

\bibitem{luongo2} P. K. S. Dunsby and O. Luongo, On the theory and applications of modern cosmography, {\it Int. J. Geom. Meth. Mod. Phys.} {\bf 13} (2016) 03, 1630002.

\bibitem{luongo3} S. Capozziello, M. De Laurentis, O. Luongo and A. C. Ruggeri, Cosmographic constraints and cosmic fluids, {\it Galaxies} {\bf 1} (2013), 216--260.

\bibitem{brans1} C. Brans and R. H. Dicke, Mach's principle and a relativistic theory of gravitation, {\it Phys. Rev.} {\bf 124} (1961), 925--935.

\bibitem{bekenstein} J. D. Bekenstein, Exact solutions of Einstein conformal scalar equations,
{\it Ann. Phys.} {\bf 82} (1974), 535--547.

\bibitem{nordt} K. Nordtvedt, PostNewtonian metric for a general class of scalar tensor gravitational theories and observational consequences, {\it Astrophys. J.} {\bf 161} (1970), 1059--1067.

\bibitem{wago} R. V. Wagoner, Scalar tensor theory and gravitational waves, {\it     Phys. Rev. D} {\bf 1} (1970), 3209--3216.

\bibitem{fujii} Y. Fujii and K. Maeda, {\it The Scalar-Tensor Theory of Gravitation}, Cambridge Monographs on Mathematical Physics {\bf 7} (Cambridge University Press, Cambridge 2003).

\bibitem{quiros} I. Quiros, Selected topics in scalar–tensor theories and beyond,     {\it Int. J. Mod. Phys. D} {\bf 28} (2019) 07, 1930012.

\bibitem{faraoni1} V. Faraoni, \textit{Cosmology in Scalar-Tensor Gravity},  Fundamental Theories of Physics Series {\bf 139}  (Springer Netherlands 2004).

\bibitem{weinberg} S. Weinberg, \textit{Gravitation and Cosmology : Principles and Applications of the General Theory of Relativity} (John Wiley and Sons, New York 1972).

\bibitem{soma} N. Banerjee and S. Sen, Does Brans-Dicke theory always yield general relativity in the infinite $\omega$ limit?, {\it Phys. Rev. D} {\bf 56} (1997), 1334--1337.

\bibitem{valerio2} V. Faraoni, Illusions of general relativity in Brans-Dicke gravity, {\it Phys. Rev. D} {\bf 59} (1999), 084021.

\bibitem{johri} C. Mathiazhagan and V. B. Johri, An inflationary universe in Brans-Dicke theory: a hopeful sign of theoretical estimation of the gravitational constant,  {\it Class. Quant. Grav.} {\bf 1} (1984), L29--L32.

\bibitem{stein} D. La and P. J. Steinhardt, Extended inflationary cosmology, {\it Phys. Rev. Lett.} {\bf 62} (1989), 376.

\bibitem{banerjee} N. Banerjee and D. Pavon, Cosmic acceleration without quintessence,  {\it Phys. Rev. D} {\bf 63} (2001), 043504. 

\bibitem{boch} N. M. Bocharova, K. A. Bronnikov, and V. N. Mel’nikov, An exact solution of
the system of Einstein equations and mass-free scalar field, {\it Vestnik Moskov.
Univ. Fizika} {\bf 25} (1970), 706–709.

\bibitem{beken2} J. D. Bekenstein, {Black holes with scalar charge}, {\it Ann. Phys.} {\bf 91} (1975), 75–82.

\bibitem{sud} D. Sudarsky and T. Zannias, Spherical black holes cannot support scalar
hair, {\it Phys. Rev. D} {\bf 58} (1998), 087502.

\bibitem{bronni} K. A. Bronnikov and Y. N. Kireyev, Instability of black holes with scalar
charge, {\it Phys. Lett.} {\bf 67A} (1978), 95–96.

\bibitem{beken3} J. D. Bekenstein, Black hole hair: twenty-five years after, in: I. M. Dremin
and A. M. Semikhatov (eds.): {\it Proceedings of the Second International Sakharov Conference on Physics, Moscow, Russia, 20-23 May 1996},  216–219 (World
Scientific, Singapore 1997).

\bibitem{xan} B. C. Xanthopoulos and T. Zannias, The uniqueness of the Bekenstein black
hole, {\it J. Math. Phys.} {\bf 32} (1991), 1875–1880.



\bibitem{win} E. Winstanley, On the existence of conformally coupled scalar field hair for black holes in (anti-)de Sitter space, {\it  Found. Phys.} {\bf 33} (2003), 111--143.

\bibitem{her} C. A. R. Herdeiro and E. Radu, Asymptotically flat black holes with scalar hair: a review,   {\it Int. J. Mod. Phys. D} {\bf 24} (2015) 09, 1542014.

\bibitem{avijit} A. Chowdhury and N. Banerjee, Quasinormal modes of a charged spherical black hole with scalar hair for 
scalar and Dirac perturbations, {\it Eur. Phys. J. C} {\bf 78} (2018) 7, 594.

\bibitem{abreu} J. P. Abreu, P. Crawford and J. P. Mimoso, Exact conformal scalar field cosmologies, {\it Class. Quant. Grav.} {\bf 11} (1994), 1919--1940.



\bibitem{thesis} C.Brans, Ph.D. thesis, Princeton University, Princeton, New Jersey (1961).

\bibitem{brans2} C. H. Brans, Mach's Principle and a Relativistic Theory of Gravitation. II, {\it Phys. Rev.} {\bf 125} (1962), 2194--2201.

\bibitem{bhadra} A. Bhadra and K. Sarkar, On static spherically symmetric solutions of the vacuum Brans-Dicke theory, {\it Gen. Relativ. Gravit.} {\bf 37} (2005), 2189-2199.

\bibitem{janis} A. I. Janis, E. T. Newman and J. Winnicour, Reality of the Schwarzschild Singularity, {\it Phys. Rev. Lett.} {\bf 20} (1968), 878-880.

\bibitem{camp} M. Campanelli and C. O. Lousto, Are black holes in Brans-Dicke theory precisely the same as a general relativity?, {\it Int. J. Mod. Phys. D} {\bf 2} (1993), 451-462.






\end{thebibliography}
\end{document}